\documentclass{revtex4}
\usepackage{amssymb}
\usepackage{amsfonts}
\usepackage{amsmath}

\input{tcilatex}

\begin{document}

\title{PDM Klein-Gordon particles in G\"{o}del-type Som-Raychaudhuri cosmic
string spacetime background}
\author{Omar Mustafa}
\email{omar.mustafa@emu.edu.tr}
\affiliation{Department of Physics, Eastern Mediterranean University, G. Magusa, north
Cyprus, Mersin 10 - Turkey.}

\begin{abstract}
\textbf{Abstract:}\ In G\"{o}del-type Som-Raychaudhuri (SR) cosmic string
spacetime background, we re-cycle the Klein-Gordon (KG) oscillators and
report their \emph{correct} exact solutions. We argue that the mathematical
collapse of the KG-equation into the 2-dimensional radial Schr\"{o}%
dinger-like oscillator does not yield that the parametric characterizations
of one is inherited by the other. The angular frequency (positive) in the
Schr\"{o}dinger case is replaced by an \emph{irrational} frequency-like
(positive and negative) in the KG-case. Inheriting the Schr\"{o}dinger
oscillator's parametric characterizations implies that at least half of the
spectra (the negative part) is lost in the process. We also introduce
KG-oscillators in pseudo-G\"{o}del SR-type spacetime that admit invariance
and isospectrality with those in G\"{o}del SR-type spacetime background. We
introduce position-dependent mass (PDM) settings to KG-particles in 4-vector
and scalar Lorentz potentials in magnetic field in the G\"{o}del SR-type
spacetime background. Four illustrative examples of fundamental nature are
discussed and their exact or conditionally exact solvability are reported.
Amongst are, a PDM KG-Coulombic particle in 4-vector and scalar Coulombic
Lorentz potentials in magnetic field, a PDM KG-Coulombic particle in
4-vector and scalar Coulombic Lorentz potentials in magnetic field at zero
vorticity, a PDM KG-Coulombic particle in equally mixed 4-vector and scalar
Coulombic Lorentz potentials in magnetic field, and a quasi-free PDM
KG-oscillator. We also emphasis that the biconfluent Heun polynomial
approach, to the effective oscillator plus Cornell type potential, yields
conditionally exact solvability that paralyzes the solution from collapsing
into that of pure Coulombic one.


\end{abstract}

\maketitle

\section{Introduction}

Early universe theories have predicted several topological structural
defects in spacetime that have inspired intensive research studies in
quantum gravity, such as domain wall \cite{Vilenkin 1983,Vilenkin 1985},
cosmic string \cite{Kibble 1976,Hiscock 1985}, global monopole \cite%
{Barriola 1989} and textures \cite{Cruz 2007}. It has been observed that the
energy levels of relativistic and non-relativistic quantum systems are
infected by such topological defects, not only in general relativity but
also in the geometrical theory of topological defects in condensed matter.
For example, in general relativity the G\"{o}del spacetime metric \cite%
{Godel 1949}, with an embedded cosmic string, introduces itself as the first
cosmological solution with rotating matter. Its compact form allowed
analytical research studies of many physical and mathematical systems in
gravitational backgrounds with rotation and causality violation. Rebou\c{c}%
as and Tiomno \cite{Tiomno 1983} have examined the conditions for spacetime
homogeneity of the Riemannian manifold with a G\"{o}del spacetime background
and found that the G\"{o}del universe is homogeneous in spacetime
(ST-homogeneous). They have shown that all ST-homogeneous G\"{o}del-type
metrics characterized by the vorticity $\Omega $ and a given value of the
parameter $\tilde{\mu}$ ($-\infty \leq \tilde{\mu}^{2}\leq \infty $) can be
transformed in polar coordinates \cite{Drukker 2004,Carvalho 2014,Tiomno1
1992,Wang 2015} to%
\begin{equation}
ds^{2}=-\left( dt+\alpha \Omega \frac{\sinh \left( \tilde{\mu}\,r\right) }{%
\tilde{\mu}^{2}}d\varphi \right) ^{2}+\alpha ^{2}\frac{\sinh \left( 2\,%
\tilde{\mu}\,r\right) ^{2}}{4\,\tilde{\mu}^{2}}d\varphi ^{2}+dr^{2}+dz^{2}.
\label{Godel metric}
\end{equation}%
Where for $\tilde{\mu}^{2}<0$ there is an infinite number of successive
causal and noncausal regions, for $0\leq \tilde{\mu}^{2}<\Omega ^{2}$ there
is one noncausal region for a given $r>r_{c}$ ( $r_{c}$ is the critical
radius), and $\tilde{\mu}=\Omega $ corresponds to the G\"{o}del spacetime
metric (the first exact G\"{o}del solution of the Einstein equation
describing a complete causal, ST-homogeneous rotating universe \cite{Tiomno
1983}). Nevertheless, at the limit $\tilde{\mu}\longrightarrow 0$ of the G%
\"{o}del spacetime metric (\ref{Godel metric}) we obtain the ST-homogeneous
Som-Raychaudhuri (SR) solution 
\begin{equation}
ds^{2}=-\left( dt+\alpha \,\Omega \,r^{2}d\varphi \right) ^{2}+\alpha
^{2}\,r^{2}d\varphi ^{2}+dr^{2}+dz^{2}  \label{Som metric}
\end{equation}%
of the Einstein field equations \cite{Som 1968}.\ Where the disclination
parameter $\alpha $ admits the values $0<\alpha <1$ in general relativity
for cosmic strings with positive curvature, $\alpha >1$ in the geometric
theory of defects in condensed matter for a negative curvature, and $\alpha
=1$ corresponds to Minkowski flat spacetime metric. Moreover, the covariant
metric tensor associated with the Som-Raychaudhuri spacetime is given by%
\begin{equation}
g_{\mu \nu }=\left( 
\begin{tabular}{cccc}
$-1\smallskip $ & $\,0\,$ & $-\alpha \Omega r^{2}$ & $\,0$ \\ 
$0$ & $1\smallskip $ & $0$ & $0$ \\ 
$-\alpha \Omega r^{2}$ & $\,0$ & $\,\alpha ^{2}r^{2}\left( 1-\Omega
^{2}r^{2}\right) \,$ & $0$ \\ 
$0$ & $0$ & $0$ & $1$%
\end{tabular}%
\right) \Longleftrightarrow g^{\mu \nu }=\left( 
\begin{tabular}{cccc}
$\left( \Omega ^{2}r^{2}-1\smallskip \right) $ & $\,0\,$ & $-\frac{\Omega }{%
\alpha }$ & $0$ \\ 
$0$ & $\,1\smallskip $ & $0$ & $0$ \\ 
$-\frac{\Omega }{\alpha }$ & $\,0$ & $\,\frac{1}{\alpha ^{2}r^{2}}\,$ & $0$
\\ 
$0$ & $0$ & $0\,$ & $\,1$%
\end{tabular}%
\right) \text{ };\text{ \ }\det \left( g\right) =-\alpha ^{2}r^{2}.
\label{g metric0}
\end{equation}%
The G\"{o}del SR- type ST-homogeneous metric (\ref{Som metric}) is the core
of the current methodical proposal.

On the other hand, effective position-dependent mass (PDM) settings in the
von Roos Schr\"{o}dinger Hamiltonian \cite{von Roos} have attracted research
interest over the last few decads both in quantum and classical mechanics
(e.g., \cite{M-L 1974,Carinena Ranada Sant 2004,Mustafa 2019,Mustafa
arXiv,Mustafa Phys.Scr. 2020, Mustafa Habib 2007,Khlevniuk 2018,Mustafa
2015,dos Santos 2021,Nabulsi1 2020,Quesne 2015,Tiwari 2013}). Hereby, it has
been asserted that the PDM notion is a metaphoric manifestation of
coordinate transformation/deformation \cite{Mustafa Algadhi 2019,Mustafa
2020,Mustafa 2022,Mustafa2 2021,Mustafa3 2021,Mustafa Algadhi1 2020}. This
would, in turn, effectively change the canonical momentum for PDM classical
systems and the momentum operator for PDM quantum systems. Namely, the PDM
canonical momentum reads $p\left( x\right) =m\left( x\right) \dot{x}$ and
consequently negative the gradient of the potential force field is no longer
given by the time derivative of the canonical momentum but rather related to
the time derivative of the pseudo-momentum (i.e., Noether momentum) $\pi
\left( x\right) =\sqrt{m\left( x\right) }\dot{x}$ \cite{Mustafa arXiv}. Yet,
in quantum mechanics the PDM-momentum operator is shown to be given by $\hat{%
p}_{j}\left( \mathbf{r}\right) =-i\left( \partial _{j}-\partial _{j}m\left( 
\mathbf{r}\right) /4m\left( \mathbf{r}\right) \right) ;\,j=1,2,3,$ and
consequently yields, in its most simplistic one-dimensional form, the von
Roos \cite{von Roos} PDM kinetic energy operator $\hat{T}=\left( \hat{p}%
_{x}\left( x\right) /\sqrt{m\left( x\right) }\right) ^{2}=-m\left( x\right)
^{-1/4}\partial _{x}m\left( x\right) ^{-1/2}\partial _{x}m\left( x\right)
^{-1/4}$, which is known in the literature as Mustafa and Mazharimousavi's 
\cite{Mustafa Habib 2007} ordering. Basically, in short, a quantum
mechanical PDM particle should be associated with the PDM-momentum operator
both for relativistic and non-relativistic PDM-quantum particles. Therefore,
in the relativistic Dirac and/or Klein-Gordon (KG) equations, the assumption
that the rest mass energy term $m\longrightarrow m+m\left( x\right) +S\left(
x\right) $ does not introduce PDM relativistic particles but rather
introduces an amendment to the already existing Lorentz scalar potential,
i.e., $\tilde{S}\left( x\right) =m\left( x\right) +S\left( x\right) $,
(c.f., e.g., \cite{ikot 2016,Ghabab 2016,Mustafa 2008,Mustafa 2007,Vitoria
2016} and references cited therein). Very recently, however, we have
introduced and discussed PDM relativistic particles \cite{Mustafa 2022} in G%
\"{u}rses spacetime background \cite{Gurses 1994}, in cosmic string
spacetime in Kaluza-Klein theory \cite{Mustafa3 2021} and in a cosmic
spacetime with a cosmic space-like dislocation \cite{Mustafa2 2021}. It
would be, therefore, interesting to study PDM KG-particles in the G\"{o}%
del-type Som-Raychaudhuri (SR) spacetime background \cite{Tiomno1 1992,Bruce
1993,Das 2008,Carvalho 2016,Garcia 2017,Ahmed1 2018,Lutfuoglu 2020}.

The organization of our paper is in order. In section 2, we recycle and
discuss a KG-oscillators in the G\"{o}del SR-type cosmic string spacetime
background. Hereby, we pinpoint and emphasis (along with the analysis
carried out by Fernandez \cite{Fernandez 2021} on the related references
cited therein, some of which are cited here) that the KG-oscillators and the
Schr\"{o}dinger-oscillator are two different quantum mechanical systems and
should not be confused with each other. The mathematical collapse of the KG
equation into Schr\"{o}dinger-like oscillator does not mean that the
parametric characterizations are copied from one into the other. By
recycling this problem, we provide solid/correct interpretation grounds to
be used in the subsequent sections. We report, in section 3, a set of
KG-oscillators in pseudo-G\"{o}del SR-type cosmic string spacetime that
admit invariance and isospectrality with the KG-oscillator discussed in
section 2. We introduce, in section 4, PDM-settings (metaphorically
speaking) for KG-particles in 4-vector and/or scalar Lorentz potentials in G%
\"{o}del SR-type cosmic string spacetime background and subjected to a
magnetic field. Therein, we discuss four illustrative examples of
fundamental nature: (i) PDM-KG particles in a 4-vector and scalar Lorentz
potentials and a magnetic field in the G\"{o}del SR-type cosmic string
spacetime background, (ii) a PDM-KG Coulombic particle in 4-vector and
scalar coulombic potentials and a magnetic field in G\"{o}del SR-type
spacetime at zero vorticity, (iii) a PDM-KG Coulombic particle in equally
mixed 4-vector and scalar Coulombic Lorentz potentials and a magnetic field
in G\"{o}del SR-type spacetime, and (iv) a quasi-free PDM KG-oscillator in G%
\"{o}del SR-type cosmic string spacetime background. Our concluding remarks
are given in section 5.

\section{Klein-Gordon oscillators in the G\"{o}del SR-type cosmic string
spacetime background}

The KG-equation describing a spin-0 particle of rest mass energy $m$
(denoting $mc^{2}$ in $c=\hbar =1$ units) in the G\"{o}del SR- type
ST-homogeneous metric (\ref{Som metric}) is given by%
\begin{equation}
\frac{1}{\sqrt{-g}}\partial _{\mu }\left( \sqrt{-g}g^{\mu \nu }\partial
_{\nu }\Psi \right) =m^{2}\Psi .  \label{KG-ST eq}
\end{equation}%
Which, in a straightforward manner, would result%
\begin{equation}
\left\{ -\partial _{t}^{2}+\left( \partial _{r}^{2}+\frac{1}{r}\partial
_{r}\right) +\left( \Omega \,r\,\partial _{t}-\frac{1}{\alpha \,r}\partial
_{\varphi }\right) ^{2}+\partial _{z}^{2}-m^{2}\right\} \Psi =0.
\end{equation}%
The substitution of%
\begin{equation}
\Psi \left( t,r,\varphi ,z\right) =\exp \left( i\left[ \ell \varphi
+k_{z}z-Et\right] \right) \psi \left( r\right) =\exp \left( i\left[ \ell
\varphi +k_{z}z-Et\right] \right) \frac{R\left( r\right) }{\sqrt{r}}
\label{Psi}
\end{equation}%
would yield%
\begin{equation}
R^{\prime \prime }\left( r\right) +\left[ \lambda -\frac{\left( \tilde{\ell}%
^{2}-1/4\right) }{r^{2}}-\tilde{\omega}^{2}\,r^{2}\right] R\left( r\right)
=0,  \label{KG-SR radial eq}
\end{equation}%
where%
\begin{equation}
\lambda =E^{2}-2\,\tilde{\ell}\,\Omega \,E-k_{z}^{2}-m^{2}\text{ ; \ }\tilde{%
\ell}=\frac{\ell }{\alpha };\text{ }\tilde{\omega}^{2}=\Omega ^{2}E^{2}.
\label{KG-SR parameters}
\end{equation}%
It is obvious that this equation (\ref{KG-SR radial eq}) represents a
KG-oscillator (as a manifestation of the G\"{o}del SR-type spacetime
structure) and resembles the one-dimensional form of the two-dimensional
radially symmetric Schr\"{o}dinger-oscillator. However, the parameter $%
\tilde{\omega}=\pm \left\vert \Omega E\right\vert $ should never be confused
with the Schr\"{o}dinger-oscillator frequency for it does not admit the same
parametric characterization..Therefore, this equation would only
mathematically inherit the textbook exact eigenvalues and eigenfunctions\ of
the Schr\"{o}dinger-oscillator so that 
\begin{equation}
\lambda =2\tilde{\omega}\left( 2n_{r}+\left\vert \tilde{\ell}\right\vert
+1\right)   \label{2D lambda}
\end{equation}%
and 
\begin{equation}
R\left( r\right) \sim r^{\left\vert \tilde{\ell}\right\vert +1/2}\exp \left(
-\frac{\tilde{\omega}r^{2}}{2}\right) L_{n_{r}}^{\left\vert \tilde{\ell}%
\right\vert }\left( \tilde{\omega}r^{2}\right) ,  \label{2D R(r)}
\end{equation}%
where $L_{n_{r}}^{\left\vert \tilde{\ell}\right\vert }\left( \tilde{\omega}%
r^{2}\right) $ are the associated Laguerre polynomials. Under such settings,
one obtains%
\begin{equation}
E=\Omega \,\left( 2n_{r}+\left\vert \tilde{\ell}\right\vert +\tilde{\ell}%
+1\right) \pm \sqrt{\Omega ^{2}\,\left( 2n_{r}+\left\vert \tilde{\ell}%
\right\vert +\tilde{\ell}+1\right) ^{2}+m^{2}+k_{z}^{2}}.
\label{KG-SR energy}
\end{equation}%
We clearly observe, hereby, that the vorticity parameter $\Omega $ plays a
magnetic field-like role of lifting the degeneracies associated with the
irrational magnetic quantum number $\tilde{\ell}=\ell /\alpha =\pm
\left\vert \tilde{\ell}\right\vert $ and results in quasi Landau energy
levels. Moreover, it should be noted here that for $\Omega =0$, $k_{z}=0$
and $\alpha =1$, the rest mass energies for the one-dimensional KG-particle
are retrieved, with $k_{z}=0$, as a natural tendency of the more general
case at hand. However, when this result is compared with that of Carvalho et
al \cite{Carvalho 2014} (i.e., equation (14) in \cite{Carvalho 2014}) we
observe that the negative energies are missed in their results and should be
corrected into those reported in (\ref{KG-SR energy}) above. So should be
the case with the results reported on the linear confinement of a scalar
particle in G\"{o}del-type space-time by Vit\'{o}ria \cite{Vitoria 2018}
(their Eq. (23)) and in the related comment by Neto \cite{Francisco 2020}
(their Eq. (18)) for zero linear confinement. That would also include the
results reported by Ahmed \cite{Ahmed 2020}, where not only half of the
spectra (the negative energies) is missed but also all nodeless (i.e.,
states with $n_{r}=0$) states.

\section{KG-oscillators in pseudo-G\"{o}del SR-type cosmic string spacetime
admitting invariance and isospectrality}

Let us consider that the G\"{o}del SR-type spacetime metric $ds^{2}$ of \ (%
\ref{Som metric}) be transformed so that%
\begin{equation}
ds^{2}\longrightarrow d\,\tilde{s}^{2}=-\left( d\tilde{t}+\alpha \,\Omega \,%
\tilde{r}^{2}d\tilde{\varphi}\right) ^{2}+\alpha ^{2}\,\tilde{r}^{2}d\tilde{%
\varphi}^{2}+d\tilde{r}^{2}+d\tilde{z}^{2}  \label{pseudo Som metric}
\end{equation}%
where%
\begin{equation}
\begin{tabular}{llll}
$d\tilde{t}=dt,d\tilde{r}=\sqrt{f\left( r\right) }dr,$ & $\tilde{r}=\sqrt{%
Q\left( r\right) }r,$ & $d\tilde{\varphi}=d\varphi ,$ & $d\tilde{z}=dz$%
\end{tabular}%
,  \label{PDM PT1}
\end{equation}%
and hence%
\begin{equation}
\frac{d\tilde{r}}{dr}\Longrightarrow \sqrt{f(r)}=\sqrt{Q\left( r\right) }%
\left[ 1+\frac{Q^{\prime }\left( r\right) }{2Q\left( r\right) }r\right] ,
\label{PDM PT2}
\end{equation}%
to govern the correlation between $f\left( r\right) $ and $Q\left( r\right) $%
. Under such transformation settings, a pseudo-G\"{o}del SR-type spacetime
metric is manifestly introduced as 
\begin{equation}
d\,\tilde{s}^{2}=-\left( dt+\alpha \,\Omega \,g\left( r\right) d\varphi
\right) ^{2}+\alpha ^{2}\,g\left( r\right) d\varphi ^{2}+f\left( r\right)
dr^{2}+dz^{2}  \label{pseudo Godel metric}
\end{equation}%
Then the covariant metric tensor associated to such a pseudo-G\"{o}del
SR-type spacetime is given by%
\begin{equation}
\tilde{g}_{\mu \nu }=\left( 
\begin{tabular}{cccc}
$-1\smallskip $ & $\,0\,$ & $-\alpha \,\Omega g\left( r\right) $ & $\,0$ \\ 
$0$ & $f\left( r\right) \smallskip $ & $0$ & $0$ \\ 
$-\alpha \,\Omega \,g\left( r\right) $ & $\,0$ & $\,\alpha ^{2}\,g\left(
r\right) \left( 1-\Omega ^{2}\,g\left( r\right) \right) \,$ & $0$ \\ 
$0$ & $0$ & $0$ & $1$%
\end{tabular}%
\right) \Longleftrightarrow \tilde{g}^{\mu \nu }=\left( 
\begin{tabular}{cccc}
$\left( \Omega ^{2}\,g\left( r\right) -1\smallskip \right) $ & $\,0\,$ & $-%
\frac{\Omega }{\alpha }$ & $0$ \\ 
$0$ & $\,\frac{1\smallskip }{f\left( r\right) }$ & $0$ & $0$ \\ 
$-\frac{\Omega }{\alpha }$ & $\,0$ & $\,\frac{1}{\alpha ^{2}\,g\left(
r\right) }\,$ & $0$ \\ 
$0$ & $0$ & $0\,$ & $\,1$%
\end{tabular}%
\right) ,  \label{g metric1}
\end{equation}%
where $g\left( r\right) =Q\left( r\right) r^{2}$ and $\det \left( g\right)
=-\alpha ^{2}\,f\left( r\right) g\left( r\right) $. We may now use similar
assumption to that in (\ref{Psi}) and define%
\begin{equation}
\Psi \left( t,\tilde{r},\varphi ,z\right) =\exp \left( i\left[ \ell \varphi
+k_{z}z-Et\right] \right) \frac{R\left( \tilde{r}\right) }{\sqrt{\tilde{r}}};%
\text{ }\tilde{r}=\sqrt{g\left( r\right) }=\sqrt{Q\left( r\right) }r.
\label{Psi1}
\end{equation}%
Consequently, the KG-equation (\ref{KG-ST eq}) would read%
\begin{equation}
\left[ \frac{1}{\sqrt{g\left( r\right) f\left( r\right) }}\partial
_{r}\left( \sqrt{\frac{g\left( r\right) }{f\left( r\right) }}\partial
_{r}\right) +\lambda -\frac{\tilde{\ell}^{2}}{g\left( r\right) }-\tilde{%
\omega}^{2}g\left( r\right) \right] \frac{R\left( \tilde{r}\right) }{\sqrt{%
\tilde{r}}}=0,
\end{equation}%
to imply, with $\tilde{r}=\sqrt{g\left( r\right) }=\sqrt{Q\left( r\right) }r$%
, that%
\begin{equation}
\frac{d^{2}R\left( \tilde{r}\right) }{d\tilde{r}^{2}}+\left[ \lambda -\frac{%
\left( \tilde{\ell}^{2}-1/4\right) }{\tilde{r}^{2}}-\tilde{\omega}^{2}\,%
\tilde{r}^{2}\right] R\left( \tilde{r}\right) =0.  \label{KG-SR radial eq1}
\end{equation}%
Obviously, this equation is invariant with (\ref{KG-SR radial eq}) and
shares the same eigenvalues. We may now conclude that the two equations (\ref%
{KG-SR radial eq}) and (\ref{KG-SR radial eq1}) are invariant and
isospectral. Moreover, our transformed metric in (\ref{pseudo Godel metric})
is a pseudo-G\"{o}del SR-type spacetime metric under our transformation
recipe in (\ref{PDM PT1}) and (\ref{PDM PT2}). This would in effect
introduce new pseudo-G\"{o}del SR-type solutions provided that $Q\left(
r\right) $ and $f\left( r\right) $ are correlated by (\ref{PDM PT2}). Yet,
one should notice that for $Q\left( r\right) =1\Longleftrightarrow $ $%
f\left( r\right) =1$, our pseudo-G\"{o}del SR-type spacetime metric (\ref%
{pseudo Godel metric}) collapses into that of G\"{o}del SR-type in (\ref{Som
metric}). This would in effect introduce a new set of the G\"{o}del SR-type
spacetime metrics that may very well be classified as pseudo-G\"{o}del
SR-type spacetime metrics that can be transformed back into the G\"{o}del
SR-type spacetime metrics through the transformation in (\ref{PDM PT2}).

\section{PDM-KG particles in a 4-vector and scalar Lorentz potentials and a
magnetic field in G\"{o}del SR-type cosmic string spacetime background}

The KG-equation for a spin-0 particle in a 4-vector potential $A_{\mu }$ in
the G\"{o}del SR-type spacetime background (\ref{Som metric})\ is given by%
\begin{equation}
\frac{1}{\sqrt{-g}}D_{\mu }\left( \sqrt{-g}g^{\mu \nu }D_{\nu }\Psi \right)
=m^{2}\Psi ,  \label{EM-KG eq}
\end{equation}%
where the gauge-covariant derivative is given by $D_{\mu }=\partial _{\mu
}-ieA_{\mu }$. Moreover, in a recent paper, we have asserted that the
PDM-setting is a manifestation of coordinate deformation/transformation \cite%
{Mustafa Algadhi 2019,Mustafa 2020,Mustafa 2022,Mustafa2 2021,Mustafa3 2021}
that yields to the introduction of the PDM-momentum operator%
\begin{equation}
\mathbf{\hat{p}}\left( \mathbf{r}\right) =-i\left( \mathbf{\nabla -}\frac{%
\mathbf{\nabla }f\left( \mathbf{r}\right) }{4f\left( \mathbf{r}\right) }%
\right) \Longleftrightarrow p_{j}=-i\left( \partial _{j}-\frac{\partial
_{j}f\left( \mathbf{r}\right) }{4f\left( \mathbf{r}\right) }\right) .
\label{PDM-op}
\end{equation}%
This would, in effect, suggest that a PDM-quantum particle (metaphorically
speaking) is described by the PDM-momentum operator in (\ref{PDM-op}) and
may be subjected to some interaction potential force fields. However, should
one be interested in invariant and isospectral systems, then our recipe in
the preceding section shall be followed (c.f., e.g., \cite{Mustafa 2020}).
In the current section, however, we shall assume that we have only a
deformation/defect in the radial coordinate so that%
\begin{equation}
\partial _{\mu }\longrightarrow \partial _{\mu }+\mathcal{F}_{\mu };\mathcal{%
F}_{\mu }=\left( 0,\mathcal{F}_{r},0,0\right) ,\mathcal{F}_{r}=\frac{%
f^{\prime }\left( r\right) }{4f\left( r\right) }  \label{f(r)}
\end{equation}%
(e.g., \cite{Mustafa 2022,Mustafa2 2021,Mustafa3 2021}). This would in turn
allow us to cast equation (\ref{EM-KG eq}) as 
\begin{equation}
\frac{1}{\sqrt{-g}}\tilde{D}_{\mu }\left( \sqrt{-g}g^{\mu \nu }\tilde{D}%
_{\nu }\Psi \right) =\left( m+S\left( r\right) \right) ^{2}\Psi ;\;\tilde{D}%
_{\mu }=\partial _{\mu }+\mathcal{F}_{\mu }-ieA_{\mu }.  \label{EM-KG eq1}
\end{equation}%
Consequently, one obtains%
\begin{equation}
\left\{ -D_{t}^{2}+\frac{1}{r}D_{r}\left( r\left[ D_{r}-\mathcal{F}_{r}%
\right] \right) +\left( \Omega \,rD_{t}-\frac{1}{\alpha r}D_{\varphi
}\right) ^{2}+D_{z}^{2}+\mathcal{F}_{r}\left( D_{r}-\mathcal{F}_{r}\right)
-\left( m+S(r)\right) ^{2}\right\} \Psi =0.  \label{General-PDM-KG}
\end{equation}%
At this point, we may now define the corresponding gauge-covariant
derivatives so that%
\begin{equation}
\begin{tabular}{llll}
$D_{t}=\partial _{t}-ieA_{t}=\partial _{t}-iV\left( r\right) ,$ & $%
D_{r}=\partial _{r},$ & $D_{\varphi }=\partial _{\varphi }-ieA_{\varphi },$
& $D_{z}=\partial _{z},$%
\end{tabular}
\label{covariant derivatives}
\end{equation}%
where, $V\left( r\right) =eA_{t}$ is the Lorentz 4-vector potential (i.e.,
transforms within the 4-vector potential $A_{\mu }$), $S\left( r\right) $ is
the Lorentz scalar potential (i.e., transforms like the rest mass energy $%
m\longrightarrow m+S\left( r\right) $), and $eA_{\varphi }$ may include both
magnetic and Aharonov-Bohm flux fields effects \cite{Mustafa3 2021,Mustafa
Algadhi1 2020}. We may now use the assumption of (\ref{Psi}) in (\ref%
{General-PDM-KG}) and obtain%
\begin{equation}
\left\{ \partial _{r}^{2}+\frac{1}{4r^{2}}+\left( E+V\left( r\right) \right)
^{2}-\left[ \Omega r\left( E+V\left( r\right) \right) +\frac{\left( \ell
-eA_{\varphi }\right) }{\alpha r}\right] ^{2}-M\left( r\right)
-k_{z}^{2}-\left( m+S(r)\right) ^{2}\right\} R\left( r\right) =0,
\label{PDM-KG1}
\end{equation}%
where%
\begin{equation}
M\left( r\right) =-\frac{3}{16}\left( \frac{f^{\prime }\left( r\right) }{%
f\left( r\right) }\right) ^{2}+\frac{1}{4}\frac{f^{\prime \prime }\left(
r\right) }{f\left( r\right) }+\frac{f^{\prime }\left( r\right) }{4rf\left(
r\right) }.  \label{M(r)}
\end{equation}

This result would describe KG-particles in G\"{o}del SR-type cosmic string
spacetime background within effective PDM introduced as a topological defect
in the coordinate system. In the following subsections, we discuss some
illustrative examples of fundamental nature.

\subsection{PDM-KG Coulomb particle in 4-vector and scalar Coulombic
potentials and a magnetic field in G\"{o}del SR-type cosmic string spacetime
background}

Let us consider a PDM-KG particle with an effective PDM function (a
dimensionless scalar multiplier) given by 
\begin{equation}
f\left( r\right) =\exp \left( 4\beta r\right) ,  \label{f(r)-Coulomb}
\end{equation}%
so that equations (\ref{f(r)}) and (\ref{M(r)}) yield%
\begin{equation}
\mathcal{F}_{r}=\frac{f^{\prime }\left( r\right) }{4f\left( r\right) }=\beta
\Longleftrightarrow M\left( r\right) =\beta ^{2}+\frac{\beta }{r}.
\label{M(r)-Coulomb}
\end{equation}%
Moreover, for $V\left( r\right) =\beta _{1}/r$, $S\left( r\right) =\beta
_{2}/r$, and $A_{\varphi }=B_{\circ }r/2$ equation (\ref{PDM-KG1}) would read%
\begin{equation}
\left\{ \partial _{r}^{2}-\frac{\left( \mathcal{L}^{2}-1/4\right) }{r^{2}}-%
\tilde{\omega}^{2}r^{2}-2E\Omega \left( \Omega \beta _{1}+\tilde{B}\right) r+%
\frac{\tilde{\beta}}{r}+\tilde{\lambda}\right\} R\left( r\right) =0,
\label{KG-Coulomb}
\end{equation}%
where%
\begin{equation}
\tilde{\lambda}=E^{2}-k_{z}^{2}-m^{2}-\tilde{B}^{2}-\Omega ^{2}\beta
_{1}^{2}-\beta ^{2}-2\,\tilde{\ell}\,\Omega \,E+2\Omega \tilde{B}\beta _{1}%
\text{ ; \ }\tilde{\ell}=\frac{\ell }{\alpha },\tilde{B}=\frac{eB_{\circ }}{%
2\alpha },  \label{lambdat}
\end{equation}%
and%
\begin{equation}
\tilde{\beta}=-\beta -2m\beta _{2}+2E\beta _{1}+2\tilde{\ell}\tilde{B}-2\,%
\tilde{\ell}\,\Omega \beta _{1},\;\mathcal{L}^{2}=\tilde{\ell}^{2}-\beta
_{1}^{2}+\beta _{2}^{2},\;\tilde{\omega}^{2}=\Omega ^{2}E^{2}.  \label{betas}
\end{equation}%
The differential equation (\ref{KG-Coulomb}) at hand admits a solution in
the form of biconfluent Heun functions so that%
\begin{equation}
\psi \left( r\right) =\frac{R\left( r\right) }{\sqrt{r}}=\mathcal{N}%
\,r^{\left\vert \mathcal{L}\right\vert }\,e^{-\left( \frac{\tilde{\omega}%
r^{2}}{2}+\Omega \beta _{1}r+\tilde{B}r\right) }\,H_{B}\left( \alpha
^{\prime },\beta ^{\prime },\gamma ^{\prime },\delta ^{\prime },\sqrt{\tilde{%
\omega}}r\right) ,  \label{PDM-KG Coulomb psi(r)}
\end{equation}%
where%
\begin{equation}
\alpha ^{\prime }=2\left\vert \mathcal{L}\right\vert \,,\,\beta ^{\prime }=%
\frac{2\left( \Omega \beta _{1}+\tilde{B}\right) }{\sqrt{\tilde{\omega}}}%
\,,\,\gamma ^{\prime }=\frac{\tilde{\lambda}+\left( \Omega \beta _{1}+\tilde{%
B}\right) ^{2}}{\tilde{\omega}}\,,\delta ^{\prime }=-\frac{2\tilde{\beta}}{%
\sqrt{\tilde{\omega}}}.  \label{PDM-KG parameters}
\end{equation}%
This would result in a biconfluent Heun polynomial of degree $n^{\prime
}=2n_{r}\geq 0$ (where $n_{r}$ denotes the radial quantum number) when $%
\gamma ^{\prime }=2\left( 2n_{r}+1\right) +\alpha ^{\prime }$ and
consequently yields%
\begin{equation}
E=\,\Omega \,\left( 2n_{r}+\left\vert \mathcal{L}\right\vert +\,\tilde{\ell}%
\,+1)\right) \pm \sqrt{\,\Omega ^{2}\,\left( 2n_{r}+\left\vert \mathcal{L}%
\right\vert +\,\tilde{\ell}\,+1)\right) ^{2}+m^{2}+k_{z}^{2}+\beta
^{2}-4\Omega \beta _{1}\tilde{B}}.  \label{PDM-KG Coulomb E}
\end{equation}%
This condition would in turn classify the solution as a conditionally exact
solution, therefore\ (c.f., e.g., \cite{Fernandez 2021,Mustafa 2022,Mustafa3
2021} for more details). Nevertheless, the choice, $n^{\prime }=2n_{r}\geq 0$%
, is manifested by the requirement that the eigenvalues as well as the
eigenfunctions should naturally retrieve those of the KG-oscillator when $%
\beta =\beta _{1}=\beta _{2}=\tilde{B}=0$ (e.g., \cite{Mustafa 2022}).
However, one would notice that for the case where the vorticity $\Omega =0$,
the differential equation in (\ref{KG-Coulomb}) effectively reduces to that
of the radial Schr\"{o}dinger Coulomb problem. However, neither the reported
biconfluent Heun polynomial in (\ref{PDM-KG Coulomb psi(r)}) nor the
eigenvalues in (\ref{PDM-KG Coulomb E}) collapse into the corresponding
known\ Schr\"{o}dinger-like KG-Coulombic solution. One has therefore to
restart from the beginning and work out the related solution as in the
sequel subsection.

In fact, Fern\'{a}ndez \cite{Fernandez 2021} has provided through and
comprehensive details on the conditional exact and/or numerical exact
solvability of the corresponding one-dimensional Schr\"{o}dinger model%
\begin{equation}
\left\{ \partial _{r}^{2}-\frac{\left( \mathcal{L}^{2}-1/4\right) }{r^{2}}%
-Ar^{2}-Br+\frac{C}{r}+E\right\} R\left( r\right) =0.  \label{Fernandez}
\end{equation}%
Fern\'{a}ndez work, along with the arguments reported in the model above,
provide a brut-force evidence on not only the conditional exactness of the
biconfluent Heun polynomials solution but also on the lack of the correct
convergence of these polynomials into the pure Coulombic (when $A=0=B$ in (%
\ref{Fernandez})) spectral problem.

\subsection{A PDM-KG Coulombic particle in 4-vector and scalar Coulombic
potentials and magnetic field in G\"{o}del SR-type spacetime at zero
vorticity}

For such a model, we choose to work with $\Omega =0$ (no vorticity) so that
equation (\ref{KG-Coulomb}) reads%
\begin{equation}
\left\{ \partial _{r}^{2}-\frac{\left( \mathcal{L}^{2}-1/4\right) }{r^{2}}+%
\frac{\tilde{\beta}}{r}+\tilde{\lambda}\right\} R\left( r\right) =0.
\label{PDM-KG Coulomb 1}
\end{equation}%
Which resembles the two-dimensional (2D) Coulombic problem that admits exact
solution in the form of Whittaker and confluent hypergeometric functions as%
\begin{equation}
R\left( r\right) \sim \mathcal{\,}W_{M}\left( -\frac{i\tilde{\beta}}{2\sqrt{%
\tilde{\lambda}}},\left\vert \mathcal{L}\right\vert ,2i\sqrt{\tilde{\lambda}}%
r\right) \sim r^{\left\vert \mathcal{L}\right\vert +1/2}\exp \left( i\sqrt{%
\tilde{\lambda}}r\right) \,_{1}F_{1}\left( \left[ \frac{1}{2}+\left\vert 
\mathcal{L}\right\vert +\frac{i\tilde{\beta}}{2\sqrt{\tilde{\lambda}}}\right]
,\left[ 1+2\left\vert \mathcal{L}\right\vert \right] ,2i\sqrt{\tilde{\lambda}%
}r\right) .  \label{R(r) PDM-KG Coulomb 1}
\end{equation}%
Consequently, a finite confluent hypergeometric polynomial of order $n_{r}$
is obtained when%
\begin{equation}
\frac{1}{2}+\left\vert \mathcal{L}\right\vert +\frac{i\tilde{\beta}}{2\sqrt{%
\tilde{\lambda}}}=-n_{r}\Longleftrightarrow \sqrt{\tilde{\lambda}}=-\frac{i%
\tilde{\beta}}{2n_{r}+2\left\vert \mathcal{L}\right\vert +1}%
\Longleftrightarrow \tilde{\lambda}=-\frac{\tilde{\beta}^{2}}{\left(
2n_{r}+2\left\vert \mathcal{L}\right\vert +1\right) ^{2}}.
\label{lambda Coulomb 1}
\end{equation}%
In fact, $\tilde{\lambda}$ in (\ref{lambda Coulomb 1}) represents the exact
eigenvalues of the 2D-Schr\"{o}dinger Coulomb problem and the exact
eigenfunctions are given by%
\begin{equation}
\psi \left( r\right) =\frac{R\left( r\right) }{\sqrt{r}}=\mathcal{N\,}%
r^{\left\vert \mathcal{L}\right\vert }\exp \left( i\sqrt{\tilde{\lambda}}%
r\right) \,_{1}F_{1}\left( -n_{r},\left[ 1+2\left\vert \mathcal{L}%
\right\vert \right] ,\Lambda r\right) ;\,\Lambda =\frac{\tilde{\beta}}{%
2n_{r}+2\left\vert \mathcal{L}\right\vert +1},  \label{PDM-KG psi(r) 1}
\end{equation}%
where $n_{r}=0,1,2,\cdots $ is the radial quantum number.

Under such settings, equation (\ref{lambdat}) along with (\ref{lambda
Coulomb 1}) would imply, with $\tilde{n}=n_{r}+\left\vert \mathcal{L}%
\right\vert +1/2$, that%
\begin{equation}
E^{2}-k_{z}^{2}-m^{2}-\tilde{B}^{2}-\beta ^{2}=-\frac{\left( -\beta -2m\beta
_{2}+2E\beta _{1}+2\tilde{\ell}\tilde{B}\right) ^{2}}{\left( 2\tilde{n}%
\right) ^{2}}\text{,}  \label{PDM-Coulomb energy 1}
\end{equation}%
where \ $\tilde{\ell}=\ell /\alpha ,\tilde{B}=eB_{\circ }/2\alpha ,$, and $%
\mathcal{L}^{2}=\tilde{\ell}^{2}-\beta _{1}^{2}+\beta _{2}^{2}$. This result
can be simplified into%
\begin{equation}
E=\frac{-\beta _{1}\Lambda _{2}\pm \sqrt{4\tilde{n}^{2}\Lambda _{1}\left( 
\tilde{n}^{2}+\beta _{1}^{2}\right) -\tilde{n}^{2}\Lambda _{2}^{2}}}{2\left( 
\tilde{n}^{2}+\beta _{1}^{2}\right) },  \label{KG-Coulomb energy 1}
\end{equation}%
where%
\begin{equation}
\Lambda _{1}=k_{z}^{2}+m^{2}+\tilde{B}^{2}+\beta ^{2};\;\Lambda _{2}=-\beta
-2m\beta _{2}+2\tilde{\ell}\tilde{B}.  \label{Lambdas}
\end{equation}%
Obviously, the term $2\tilde{\ell}\tilde{B}$ (i.e., the magnetic field
effect) in $\Lambda _{2}$ removes the degeneracies associated with $%
\left\vert \mathcal{L}\right\vert $ in the new irrational quantum number $%
\tilde{n}=n_{r}+\left\vert \mathcal{L}\right\vert +1/2$ of (\ref{KG-Coulomb
energy 1}). Moreover, one should notice that the PDM settings of (\ref%
{f(r)-Coulomb}) and (\ref{M(r)-Coulomb}) have provided a Coulomb-like
interaction as a byproduct of its own.

\subsection{A PDM-KG Coulombic particle in equally mixed 4-vector and scalar
Lorentz potentials and a magnetic field in G\"{o}del SR-type spacetime}

We now consider a PDM particle described by (\ref{f(r)-Coulomb}) in G\"{o}%
del SR-type spacetime in a magnetic field along with equally mixed 4-vector
and scalar Coulombic Lorentz potentials and a magnetic field, where $V\left(
r\right) =S\left( r\right) =\beta _{\circ }/r:\,$ $\beta _{1}=\beta
_{2}=\beta _{\circ }$. In this case, equation (\ref{KG-Coulomb}) reads 
\begin{equation}
\left\{ \partial _{r}^{2}-\frac{\left( \,\tilde{\ell}\,^{2}-1/4\right) }{%
r^{2}}-\tilde{\omega}^{2}r^{2}-2E\Omega \left( \Omega \beta _{\circ }+\tilde{%
B}\right) r+\frac{\tilde{\beta}}{r}+\tilde{\lambda}\right\} R\left( r\right)
=0,  \label{KG-Coulomb 1}
\end{equation}%
where%
\begin{equation}
\tilde{\lambda}=E^{2}-k_{z}^{2}-m^{2}-\tilde{B}^{2}-\Omega ^{2}\beta _{\circ
}^{2}-\beta ^{2}-2\,\tilde{\ell}\,\Omega \,E+2\Omega \tilde{B}\beta _{\circ }%
\text{ ; \ }\tilde{\ell}=\frac{\ell }{\alpha },\tilde{B}=\frac{eB_{\circ }}{%
2\alpha },  \label{Coulomb 1 parameters}
\end{equation}%
and%
\begin{equation}
\tilde{\beta}=-\beta -2m\beta _{\circ }+2E\beta _{\circ }+2\tilde{\ell}%
\tilde{B}-2\,\tilde{\ell}\,\Omega \beta _{\circ },\;\tilde{\omega}%
^{2}=\Omega ^{2}E^{2}.  \label{Coulomb 1 parameters 1}
\end{equation}%
Which admits conditionally exact solution in the form of biconfluent Heun
polynomials so that%
\begin{equation}
\psi \left( r\right) =\frac{R\left( r\right) }{\sqrt{r}}=\mathcal{N}%
\,r^{\left\vert \tilde{\ell}\,\right\vert }\,e^{-\left( \frac{\tilde{\omega}%
r^{2}}{2}+\Omega \beta _{\circ }r+\tilde{B}r\right) }\,H_{B}\left( \alpha
^{\prime },\beta ^{\prime },\gamma ^{\prime },\delta ^{\prime },\sqrt{\tilde{%
\omega}}r\right) ,  \label{Coulomb 1 Heun}
\end{equation}%
where%
\begin{equation}
\alpha ^{\prime }=2\left\vert \tilde{\ell}\,\right\vert \,,\,\beta ^{\prime
}=\frac{2\left( \Omega \beta _{\circ }+\tilde{B}\right) }{\sqrt{\tilde{\omega%
}}}\,,\,\gamma ^{\prime }=\frac{\tilde{\lambda}+\left( \Omega \beta _{\circ
}+\tilde{B}\right) ^{2}}{\tilde{\omega}}\,,\delta ^{\prime }=-\frac{2\tilde{%
\beta}}{\sqrt{\tilde{\omega}}},  \label{Heun 1 parameters}
\end{equation}%
with the condition $\gamma ^{\prime }=2\left( 2n_{r}+1\right) +\alpha
^{\prime }$ to consequently yield%
\begin{equation}
E=\,\,\Omega \left( 2n_{r}+\left\vert \tilde{\ell}\right\vert +\,\tilde{\ell}%
\,+1\right) \pm \sqrt{\Omega ^{2}\left( 2n_{r}+\left\vert \tilde{\ell}%
\right\vert +\,\tilde{\ell}\,+1\right) ^{2}+m^{2}+k_{z}^{2}+\beta
^{2}-4\Omega \beta _{\circ }\tilde{B}}.  \label{Coulomb 1 energy}
\end{equation}

In each of the three PDM-KG Coulombic particles above, we notice that the
first term of the energy eigenvalues (i.e., equations (\ref{PDM-KG Coulomb E}%
),(\ref{KG-Coulomb energy 1}), and (\ref{Coulomb 1 energy})) lifts the
degeneracies associated with the magnetic quantum number $\ell =\pm
\left\vert \ell \right\vert $.

\subsection{A quasi-free PDM KG-oscillator in G\"{o}del SR-type cosmic
string spacetime background}

We now consider a free PDM KG-particle with not only position-dependent but
also energy-dependent scalar multiplier deformation function in the form of 
\begin{equation}
f\left( r\right) =\exp \left( 2\xi Er^{2}\right) ,  \label{PDM KG-oscillator}
\end{equation}%
which implies that%
\begin{equation}
M\left( r\right) =E^{2}\xi ^{2}r^{2}+2E\xi .  \label{Oscillator PDM}
\end{equation}%
Let us assume that this PDM KG-particle is free from the Lorentz potentials
(i.e., $V\left( r\right) =S\left( r\right) =0$ and no magnetic field effect $%
eA_{\varphi }=0$, hence quasi-free). In this case, equation (\ref{PDM-KG1})
reads%
\begin{equation}
\left\{ \partial _{r}^{2}+\frac{\left( \tilde{\ell}\,^{2}-1/4\right) }{r^{2}}%
-\tilde{\Omega}^{2}r^{2}+{\Large \tilde{\varepsilon}}\right\} R\left(
r\right) =0,  \label{PDM HO1}
\end{equation}%
which resembles the 2D radial Schr\"{o}dinger harmonic oscillator problem
with%
\begin{equation}
{\Large \tilde{\varepsilon}=}E^{2}-2\tilde{\ell}\Omega E-2E\xi -\left(
m^{2}+k_{z}^{2}\right) ,\;\tilde{\Omega}^{2}=E^{2}\left( \Omega ^{2}+\xi
^{2}\right) .  \label{PDM HO1 parameters}
\end{equation}%
Under such settings, the textbook solution is therefore given by%
\begin{equation}
{\Large \tilde{\varepsilon}}=2\tilde{\Omega}\left( 2n_{r}+\left\vert \tilde{%
\ell}\right\vert +1\right)   \label{Et oscillator}
\end{equation}%
and 
\begin{equation}
R\left( r\right) \sim r^{\left\vert \tilde{\ell}\right\vert +1/2}\exp \left(
-\frac{\tilde{\Omega}r^{2}}{2}\right) L_{n_{r}}^{\left\vert \tilde{\ell}%
\right\vert }\left( \tilde{\Omega}r^{2}\right) ,  \label{R(r) oscillator}
\end{equation}%
where $L_{n_{r}}^{\left\vert \tilde{\ell}\right\vert }\left( \tilde{\Omega}%
r^{2}\right) $ are the associated Laguerre polynomials. One should notice
that the choice of the deformation in (\ref{PDM KG-oscillator}) is just to
facilitate the calculation to be able to get%
\begin{equation}
E=\tilde{\gamma}\pm \sqrt{\tilde{\gamma}^{2}+m^{2}+k_{z}^{2}};\;\tilde{\gamma%
}=\tilde{\ell}\Omega +\xi +\left( \Omega ^{2}+\xi ^{2}\right) \left(
2n_{r}+\left\vert \tilde{\ell}\right\vert +1\right) .
\label{PDM quasi HO energies}
\end{equation}%
This particular example shows that PDM-KG particles (PDM-particles in
general) may very well create their own byproducted effective interactions
that may lead to bound states. Hence the notion of quasi-free particles is
unavoidable in the process. Moreover, we observe that quasi-Landau type
energy levels manifestly introduced by the vorticity parameter $\Omega $ in $%
\;\tilde{\gamma}$. Hence, the vorticity parameter have similar effect as
that of the magnetic field..

\section{Concluding remarks}

The harmonic oscillator and Coulomb problems form the most fundamental
systems in non-relativistic and relativistic quantum mechanics in different
spacetime backgrounds. Their exact solvability and impressive/superb
pedagogical/research implementation makes them systems of great physical
relevance. In the current methodical proposal, we considered KG-particles
(harmonic oscillator and Coulombic types) in G\"{o}del SR-type cosmic string
spacetime backgrounds and under the influence of different 4-vector and/or
scalar Lorentz potentials settings \cite{Quigg 1979}. We started with the
KG-oscillators in G\"{o}del SR-type cosmic string spacetime backgrounds and
reported their exact solutions. Hereby, we have emphasized that the
mathematical collapse of the KG-equation into the two dimensional radially
symmetric Schr\"{o}dinger-oscillator (hence the metaphoric notion of \emph{%
KG-oscillator} emerged in the process) does not mean that the parametric
characterizations are copied from one to the other. The parameter $\tilde{%
\omega}$ of (\ref{KG-SR parameters}) is not related to the angular
frequencies of the harmonic oscillator (e.g., \cite{Vitoria 2018,Francisco
2020,Ahmed 2020,Mirza 2004}) but rather admits values given by $\tilde{\omega%
}=\pm \left\vert \Omega E\right\vert $ (e.g., \cite{Fernandez 2021}). The
usage of $\tilde{\omega}$ as the angular frequency of the harmonic
oscillator had the consequences of losing at least half of the spectra (only
positive energies were reported and anti-particle solutions were dismissed
from the relativistic theory). The recycling of the KG-oscillators in G\"{o}%
del SR-type cosmic string spacetime backgrounds in section 2 is unavoidable,
therefore. We have, in the same section, pinpointed that the results
reported by Carvalho et al \cite{Carvalho 2014} on the quantum influence of
topological defects in G\"{o}del spacetime, by Vit\'{o}ria et al \cite%
{Vitoria 2018} on the linear confinement of a scalar particle in G\"{o}%
del-type spacetime (including the related comment by Neto et al \cite%
{Francisco 2020}), and those reported by Ahmed \cite{Ahmed 2020} on
KG-Coulombic type particles should be redirected to reflect the results of
the current proposal, therefore. Not only they have lost all negative
energies but also they have lost all nodeless states form the spectra
(excluding Carvalho et al \cite{Carvalho 2014}, where nodeless states, with $%
n_{r}=0$, are reported therein). Here, we have just mentioned few of so many
references that are not cited herein but lie far beyond the scope of the
current study. Subsequently, we have reported (in section 3) the
KG-oscillators in pseudo-G\"{o}del SR-type cosmic string spacetime that
admit invariance and isospectrality with the KG-oscillator.

On the PDM settings side of the current methodical proposal, we have
introduced a general recipe for PDM-KG particles in a 4-vector and scalar
Lorentz potentials and a magnetic field in G\"{o}del SR-type cosmic string
spacetime background (in section 4). \ Therein, we have considered four
illustrative examples of fundamental nature: (i) a PDM-KG Coulomb particle
in 4-vector and scalar Coulombic potentials and a magnetic field in G\"{o}%
del SR-type cosmic string spacetime background, (ii) a PDM-KG Coulombic
particle in 4-vector and scalar Coulombic potentials and magnetic field in G%
\"{o}del SR-type spacetime at zero vorticity, (iii) a PDM-KG Coulombic
particle in equally mixed 4-vector and scalar Lorentz potentials and a
magnetic field in G\"{o}del SR-type spacetime, and (iv) a quasi-free PDM
KG-oscillator in G\"{o}del SR-type cosmic string spacetime background.

In the light of our experience on the PDM KG-particles (this would also
include constant mass settings with $f\left( r\right) =1$, in general) in G%
\"{o}del SR-type cosmic string spacetime background, the following notes are
inevitably unavoidable:

\begin{enumerate}
\item For the PDM-KG Coulomb particle in 4-vector and scalar Coulombic
potentials and a magnetic field, the corresponding one-dimensional Schr\"{o}%
dinger equation (\ref{KG-Coulomb}), indulges within a harmonic
oscillator-like in a Cornel type potentials \cite{Quigg 1979}, that admits a
biconfluent Heun polynomial type solution (\ref{PDM-KG Coulomb psi(r)})
which consequently yields some conditionally exact energies reported in (\ref%
{PDM-KG Coulomb E}). However, it is obvious that neither the biconfluent
Heun polynomials in (\ref{PDM-KG Coulomb psi(r)}) nor the eigenvalues in (%
\ref{PDM-KG Coulomb E}) collapse into the corresponding known\ Schr\"{o}%
dinger-like KG-Coulombic solution, when the linear plus oscillator like
potential terms are switched off..

\item We have deliberately considered a PDM-KG Coulombic particle in G\"{o}%
del SR-type spacetime at zero vorticity to provide a brut-force evidence on
the drawback of the use of the biconfluent Heun polynomials approach. We
have observed that the more general and conditionally exact solution for
system (\ref{KG-Coulomb}) does not collapse into that for the pure Coulombic
one of (\ref{PDM-KG Coulomb 1}).

\item The equally mixed 4-vector and scalar Lorentz potentials is a common
practice in quarkonium spectroscopy \cite{Quigg 1979,Mustafa 1998}. We have
therefore introduced such feasible mixture/model of interaction potentials.
Again, this system suffers from the drawback of the conditional exact
solvability of the biconfluent Heun polynomial approach, as in point 1 above.

\item PDM-KG particles (PDM-particles in general) may generate their own
byproducted effective interactions that support bound states. They should,
therefore, be labelled as quasi-free PDM-particles. Yet, quasi-Landau type
energy levels could be introduced by the vorticity $\Omega .$of spacetime
that has similar effect of the magnetic field.
\end{enumerate}

The drawback in the biconfluent Heun polynomial approach (a power series
manifested solution) inspires the search for a more sophisticated/reliable
approach. We foresee that an alternative could be sought in the well known
shifted $1/N$ or shifted-$\ell $ expansion quasi-perturbation method (e.g., 
\cite{Imbo 1983,Imbo 1984,Mustafa 1993,Mustafa 1994,Mustafa 1991,Mustafa
1998,Mustafa 2000,Mustafa 2002}). Such quasi-perturbation methods result the
exact energies for the harmonic oscillator like and Coulomb like
interactions (especial cases of \ (\ref{PDM-KG Coulomb psi(r)})) from the
zeroth (leading) order correction, where all higher order correction
identically vanish. This method is also reported to provide highly accurate
results for the Coulomb plus oscillator \cite{Mustafa 1993,Mustafa 1994}, a
close form of potential to the one used in section 4-A and 4-C. Work in this
direction is in progress.

\end{document}